# Some Proxy Signature and Designated Verifier Signature Schemes over Braid Groups

***Sunder Lal and Vandani Verma***
***Department of Mathematics, Dr. B.R.A. (Agra), University, Agra-282002 (UP), India***
**E-mail- sunder_lal2@rediffmail.com, vandaniverma@yahoo.com**

**Abstract:** Braids groups provide an alternative to number theoretic public cryptography and can be implemented quite efficiently. The paper proposes five signature schemes: Proxy Signature, Designated Verifier, Bi-Designated Verifier, Designated Verifier Proxy Signature And Bi-Designated Verifier Proxy Signature scheme based on braid groups. We also discuss the security aspects of each of the proposed schemes.

## 1. Introduction

Braid groups have recently attracted the attention of many cryptographers. These highly non-commutative groups are useful to construct cryptosystems They provide numerous mathematical hard problems such as the conjugacy search problem, braid decomposition problem and root problem. Moreover, the group operation and generation of the parameters can be implemented quite efficiently. Braid groups were introduced in 1947 by Artin [2] and were first used to construct a Diffie-Hellmann type key agreement protocol and a public key encryption scheme by Ko et al [9]. Many cryptographic protocols based on braid groups have since been proposed by Anshel et al [1], Cha et al [3], Ko et al [11], Kim et al [8], Thomas et al [21], Girraj [22] etc.

Proxy signatures were first proposed by Mambo et al [18] in 1996. In the proxy signatures an original signer can designate a person as his/her proxy signer and delegate the power to sign the digital documents on his behalf. Depending upon the types of delegation, the proxy signature may be classified as full delegation, partial delegation or delegation by warrant. If the original signer gives his private key to the proxy signer, the proxy signer gets the same signing capability as the original signer. Such delegation is called **full delegation**. For most of the real world settings, full delegation is impractical and insecure. If the original signer generates a proxy secret key for the proxy signer, which the proxy signer uses to sign message on behalf of the original signer, but is unable to derive the original signer's secret key, the delegation is called **partial**. However, if the original signer gives the proxy signer a warrant composed of a message part, time of validation of proxy signature and public key part, the delegation is called **by warrant**. Combining partial delegation and delegation by warrant one may get **partial delegation with warrant**. Based on these concepts, several proxy signature schemes [4, 6, 7, 12, 17, 18, 20, 23] have been proposed

Jakobsson et al [5] proposed the designated verifier signature (DVS) scheme in 1996. It is a special type of signatures that provides message authentication without non-repudiation. In such signature schemes only the designated verifier can check the validity of the signatures and he cannot convince anyone else of this fact, as he himself is able to produce the indistinguishable signatures. Saeednia [19] in 2003 added the concept of strongness in DVS to make them strong designated verifier signatures (SDVS), such that the designated verifier can only verify the signatures by using his secret key during verification phase. There exist several DVS scheme in literature [10, 15, 16, 24]. The concept of proxy signatures is added to DVS to generate0 designated verifier proxy signature schemes (DVPS). Lal et al [13] proposed some DVPS signature schemes in ID based cryptography. Lal et al [14] also proposed bi-designated verifier proxy signature schemes in which the proxy signature can only be verified by the two verifiers designated by the original signer.

There are several proxy schemes with partial delegation by warrant which are based on number theoretic groups however; there exists no such scheme on braid groups. In this paper, we first propose 'Proxy Signature Scheme With Partial Delegation By Warrant' on braid groups, 'A Designated Verifier Signature' (DVS) scheme and 'Bi-Designated Verifier Signature' scheme (Bi-DVS) and then combine this proxy signature scheme to the DVS and Bi-DVS to form 'Designated Verifier Proxy Signature' (DVPS) scheme and 'Bi-Designated Verifier Proxy Signature' (Bi-DVPS) scheme. In our proxy signature schemes, a warrant containing the identities of the original signer, the proxy signer and period of delegation is attached to the signatures, such signatures are valid for a certain period of time and the verifier can check the authenticity of the proxy signatures by checking the identity of the proxy signer in the warrant. The security of our schemes is based on the conjugacy search problem of braid groups.

The rest of the paper is organized as follows: in section 2, we state some preliminaries, in section 3 we present our proxy signature scheme, in section 4 we propose designated verifier signature scheme and in section 5 we present bi-designated verifier signature scheme. In section 6 we present designated verifier proxy signature (DVPS) scheme and in section 7 we propose bi-designated verifier proxy signature (Bi-DVPS) scheme. Finally, the section 8 concludes the paper.

## 2. Preliminaries

This section discusses the basic definitions of braid groups and some hard problems on these groups.

For $n \geq 2$, the group $B_n$ of $n$-braid is the group generated by $\sigma_1, \sigma_2 \ldots \sigma_{n-1}$ with the conditions

(i) $\sigma_i \sigma_j = \sigma_j \sigma_i, where \left| i - j \right| \geq 2$

(ii) $\sigma_i \sigma_{i+1} \sigma_i = \sigma_{i+1} \sigma_i \sigma_{i+1}$

Each element of the group $B_n$ is called an $n$ – braid and the integer n is called the braid index. An $n$-braid where *'n'* is an integer is a set of disjoint $n$-strands all of which are attached to two horizontal bars at the top and at the bottom in such a way that each strands always heads downward as one follows the path along the strand from the top to the bottom. In a set of braids if one braid can be deformed to the other continuously then the two braids are said to equivalent The multiplication of two braids '*a*' and '*b*' can be obtained by positioning *'a'* on the top of *'b'*. The identity of group $B_n$ is '*e*' which consists of 'n' straight vertical strands and the reflection of a with respect to a horizontal line is the inverse of '*a*'. So by switching the over-strand and under-strand $\sigma^{-1}$can be obtained from σ.

We now describe some mathematically hard problems in braid groups. Two elements $x$ and $y$ are conjugate i.e. $x \sim y$ if there is an element *'a'* such that $y = axa^{-1}$. For $m < n$, $B_m$ can be considered as a subgroup of $B_n$ generated by $\sigma_1, \sigma_2 \ldots \sigma_{n-1}$. In each of the schemes $H_1: \{0, 1\}^* \rightarrow B_{l+r}$ and $H_2: B_{l+r} \rightarrow \{0, 1\}^*$ are the one way hash functions.

- **Conjugacy Decision Problem (CDP)**
  Given $(x, y) \in B_n \times B_n$. Determine whether $x$ and $y$ are conjugate.
- **Conjugacy Search Problem (CSP)**
  Given $(x, y) \in B_n \times B_n$ which are conjugates find $b \in B_n$ such that $y = bxb^{-1}$
- **Generalized Conjugacy Search Problem (GCSP)**
  Given $(x, y) \in B_n \times B_n$ such that $y = axa^{-1}$ for some $a \in B_m$ ; $m \leq n$. find $b \in B_n$ such that $y = bxb^{-1}$

- **Conjugacy Decomposition Problem (CDP)**
  Given $(x, y) \in B_n \times B_n$ such that $y = axa^{-1}$ for some $a \in B_n$, $m < n$. Find $b_1, b_2 \in B_n$ such that $y = b_1 x b_2$.

The public key system on braid groups is based on the generalized conjugacy search problem. We consider two subgroups $LB_l$ and $RB_r$ of $B_{l+r}$ for some appropriate pair of integers $(l, r)$. $LB_l$ (resp. $RB_r$) is the subgroup of $B_{l+r}$ consisting of braids made by braiding left $l$ (resp. right $r$)-strands among $(l + r)$ strands. $LB_l$ is generated by $\sigma_1, \sigma_2 \ldots \sigma_{l-1}$ and $RB_r$ is generated by $\sigma_{l+1}, \ldots \sigma_{l+r-1}$ For any $a \in LB_l$ and $b \in RB_r$, $ab = ba$.

## 3. Proposed Proxy Signature Scheme with Delegation by Warrant

In this section we propose our proxy signature scheme with delegation by warrant. Let '*m*' be the message to be signed, $m_w$ is the warrant on message '*m*' consisting of the identities of the original signer Alice, the proxy signer Bob and the period of delegation,

- **Key Generation:** Each user '*u*' chooses a braids $x_u \in_R B_{l+r}$ s.t $x_u \in_R LB_l$ and chooses $(x'_u = a_u x_u a_u^{-1}, a_u) \in_R RB_r$ such that $a_u$ is the secret key and $(x'_u, x_u)$ is the public key of the user.

- **Proxy Key Generation:** Original signer chooses a message '*m*', a warrant '$m_w$' a braid $z_o \in LB_l$ and computes $t_o = a_o z_o a_o^{-1}$ and sends $(m_w, z_o, t_o)$ to the proxy signer Bob. Bob checks $t_o x'_o \sim z_o x_o$. If this holds then computes the proxy key $PK = a_p t_o a_p^{-1}$.

- **Proxy Signature Generation:** The proxy signer Bob chooses a braid $b \in LB_l$ and computes $h = H_1(H_2(t_o x'_o) \oplus m_w)$, $\gamma = bhb^{-1}$, $\delta = bx_p b^{-1}$, $\theta = b a_p^{-1} (PK) a_p b^{-1}$. Sends $(\gamma, \delta, \theta, t_o, m_w)$ to the verifier Cindy.

- **Proxy Signature Verification:** Verifier on receiving $(\gamma, \delta, \theta, t_o, m_w)$ checks the warrant '$m_w$' computes $h = H_1(H_2(t_o x'_o) \| m_w)$ and accepts the signature if and only if $\gamma\theta \sim h\, t_o$, $\gamma\delta \sim h\, x_p$ holds.

- **Correctness:** The following equation gives the correctness of the verification equations:

- $\gamma\,\theta = (bhb^{-1})\,(b\,a_p^{-1}\,(PK)a_pb^{-1})$

  $= bh\,(a_p^{-1}\,(PK)a_p)b^{-1}$

  $= b(h\,t_o)b^{-1}$

  i.e. $\gamma\,\theta \sim h\,t_o$

- $\gamma\,\delta = (bhb^{-1})\,(bx_pb^{-1})$

  $= b(h\,x_p)b^{-1}$

  i.e. $\gamma\,\delta \sim h\,x_p$

- **Security Analysis:** Regarding security analysis of the scheme we have the following:

- **Secrecy of the proxy key:** The signature 'σ' will not reveal the proxy key $PK = a_pt_o\,a_p^{-1}$

  **Explanation:** According to braid groups, even if the pair ($\mathbf{x}'_p$, $x_p$) and ($\mathbf{x}'_o$, $x_o$) are known to user he cannot obtain $a_p$ and $a_o$ because for given $x'_p = a_px_pa_p^{-1}$ (resp. $x'_o = a_ox_oa_o^{-1}$) finding $a_p$ and $a_o$ is a conjugacy search problem. The proxy key also involves a secret braid $z_o \in LB_l$ chosen by the original signer. So. Even if one knows the secret keys of the original signer and the proxy signer he will not be able to construct the valid proxy signatures. So, signature 'σ' will not reveal the proxy key PK.

- **Signer protection:** Only the legal signer can generate the valid proxy signatures.

  **Explanation:** The legal proxy signer can only generate a valid proxy key as the construction of proxy key $PK = a_pt_o\,a_p^{-1}$ involves the secret key '$a_p$' of the proxy signer. Moreover, the original signer restricts others user's from forgeing the signatures attaches a warrant '$m_w$' (containing the identity of proxy signer) to the signatures.

- **Proxy protection:** No one can generate the proxy signatures except for the real proxy signer.

  **Explanation:** As shown in 4.2 only the legal signer can create the valid proxy key. So, no one including the original signer can construct the valid proxy key and the valid proxy signatures.

- **Original Signer Protection:** The signer indeed authorizes the proxy signer. **Explanation:** The original signer authorizes the proxy signer to generate the proxy signatures for the 'm' through a warrant '$m_w$' that contains the identity of the proxy signer, message to signed and the period of delegation.

## 4. Proposed Single Designated Verifier Signature Scheme

This section proposes the single designated verifier signature scheme. The security of the scheme is based on the conjugacy problem and the base problem in braid groups. Here we have assumed Alice as the original signer, Bob as the proxy signer and Cindy as the designated verifier chosen by the Alice. Let 'm' be the message to be signed.

- **Key Generation:** Same as section 3.

- **Signature Generation:** The signer Alice chooses a message '*m*' and a braid $b \in LB_l$ and computes $\alpha = bx_c b^{-1}$, $\beta = b\,x'_c\,b^{-1}$, $h = H_1(H_2(\beta) \oplus m)$, $\delta = a_o h\, a_o^{-1}$ Sends $\sigma = (m, \alpha, \delta)$ to the designated verifier Cindy as the signature on message '*m*'.

- **Signature Verification:** Cindy on receiving the signatures σ computes $\beta = a_c \alpha\, a_c^{-1}$, $h = H_1(H_2(\beta) \oplus m)$ and accepts the signature if and only if $\delta \sim h$, $\delta\, x'_o \sim h x_o$ holds.

- **Correctness:** The following equations give the correctness of the verification:
  - $\delta = a_o h\, a_o^{-1}$
    i.e. $\delta \sim h$
  - $\delta\, x'_o = (a_o h\, a_o^{-1})(\, a_o x_o a_o^{-1})$
    $= a_o (h\, x_o\,)\, a_o^{-1}$
    i.e. $\delta\, x'_o \sim h x_o$

- **Applications**

**Strong Designated Verifier Signature** schemes proposed here have several practical applications in the situations where the signer wishes to convince only one person about the validity of the signatures. One application of SDVS schemes is where tenderers use SDVS to digitally sign their quotations. Another application of SDVS is in software licensing. Software companies' use digitally signed keys as there software license so that these keys can only be used by the person who buys the product. Use of SDVS to produce digitally signed keys/licenses protects illegal distribution of the software.

- **Security Analysis:**

➢ **Strongness:** Only the designated verifier can verify the signatures.

**Explanation:** Only the designated verifier can verify the signatures, non-designated verifiers cannot verify the signatures. Firstly, the designated verifier have secret key $a_c$, computering $\beta = a_c \alpha a_c^{-1}$, $h = H_1(H_2(\beta) \oplus m)$. Secondly, if non-designated verifier wants to verify the signatures they must compute $\beta$ and $h$. But they do not hold the secret key $a_c$ of the designated verifier. However, the security of finding $\beta$ is based on ***Base problem 1*** (which is mathematically difficult because of the previous cryptographic assumptions).

***Base problem 1:***

***Instance:*** *The triple* $(x_c, \alpha, x'_c)$ *of elements in* $B_{l+r}$ *such that* $\alpha = bx_c b^{-1}$ *and* $x'_c = a_c x_c a_c^{-1}$ *for some hidden* $a_c \in RB_n$ *and* $b \in LB_l$.

***Objective:*** *Find* $a_c \alpha a_c^{-1}$ $(= a_c\, bx_c a_c^{-1} b^{-1})$

Thus, non-designated verifiers cannot compute β, nor carry out the verification.

➢ **Unforgeability:** Non-designated verifiers cannot forge the signatures.

**Explanation:** Suppose an opponent captures the signatures $(m, \alpha, \delta)$ and he try to operate the forgery from the condition of the verification i.e. he wants to determine $h$ $(= H_1(H_2(\beta) \oplus m))$ such that the condition of verification $(\delta x'_o \sim hx_o,\ \delta \sim h)$ are satisfied. But for obtaining '$h$' he must compute $\beta$ $(= a_c \alpha a_c^{-1})$. But $\beta$ $(= a_c \alpha a_c^{-1})$ uses the secret key of the verifier. Hence, he cannot forge the signatures from the condition of verification.

## 5. Proposed Bi-Designated Verifier Signature Scheme

This section proposes the bi-designated verifier signature scheme. We have added the one more designated verifier (Trevor) in the scheme stated in section 4 to propose our bi-designated signature scheme and the security of the scheme is based on the base problem in braid groups. Moreover, the scheme is constructed in such a manner that the two verifiers can verify the signatures individually even if they do not know anything about each other.

- **Key Generation:** Same as section 3.

- **Signature Generation:** The signer Alice chooses a message '*m*' and a braid $b \in LB_l$ and computes $\alpha_1 = bx_c b^{-1}$, $\beta_1 = b x'_c b^{-1}$, $\alpha_2 = bx_T b^{-1}$, $\beta_2 = b x'_T b^{-1}$, $h = H_1(H_2(\beta_1\beta_2) \oplus m)$, $\delta = a_o h a_o^{-1}$. Sends $\sigma_1 = (m, \alpha_1, \beta_2, \delta)$ to the designated verifier Cindy and $\sigma_2 = (m, \alpha_2, \beta_1, \delta)$ as the signature on message '*m*' to the designated verifier Trevor.

- **Signature Verification:** Cindy on receiving the signatures $\sigma_1$ computes $\beta_1 = a_c\alpha_1 a_c^{-1}$, $h = H_1(H_2(\beta_1\beta_2) \oplus m)$, and accepts the signature if and only if $\delta \sim h$ and $\delta x'_o \sim hx_o$ holds.

  Similarly, Trevor on receiving $\sigma_2$ computes $\beta_2 = a_T\alpha_2 a_T^{-1}$, $h = H_1(H_2(\beta_1\beta_2) \oplus m)$, and accepts the signature if and only if $\delta \sim h$ and $\delta x'_o \sim hx_o$ holds.

- **Correctness:** The following equations give the correctness of the verification:
  - $\delta = a_o h a_o^{-1}$ *i.e.* $\delta \sim h$
  - $\delta x'_o = (a_o h a_o^{-1})(a_o x_o a_o^{-1})$

    $= a_o (h x_o) a_o^{-1}$

    *i.e.* $\delta x'_o \sim hx_o$

- **Applications:**

Consider the following situation: Alice is starting a new project and for certain reasons he needs to discuss the same project with both Bob and Cindy for their help and suggestions. But Alice does not want his project details to get leaked while discussion with Bob and Cindy. Moreover, he does not want Bob and Cindy to convince any other person that this project is being started by Alice. In this situation Strong Bi-Designated Verifier Signature Scheme as proposed in chapter 4, may be considered very useful. As for the discussions with Bob and Cindy, Alice will digitally sign his project proposal using SBiDVS.

- **Security Analysis:**

➢ **Strongness:** Only the designated verifiers can verify the signatures.

**Explanation:** Only the designated verifiers can verify the signatures, non-designated verifiers cannot verify the signatures. Firstly, the designated verifiers have secret key $a_c$ and $a_T$, computering $\beta_1 = a_c\alpha_1 a_c^{-1}$, $\beta_2 = a_T\alpha_2 a_T^{-1}$ and $h = H_1\ (H_2(\beta_1\beta_2) \oplus m)$ . Secondly, if non-designated verifiers want to verify the signatures they must compute $\beta_1$ and $\beta_2$. But they do not hold the secret key $a_c$ and $a_T$ of the designated verifiers. However, the security of finding $\beta_1$ and $\beta_2$ is also based on **Base problem 1**

Thus, non-designated verifiers cannot compute $\beta_1$ and $\beta_2$, nor carry out the verification.

- **Unforgeability:** Non-designated verifiers cannot forge the signatures.
  **Explanation:** Suppose an opponent captures the signatures $(m,\ \alpha_1,\ \beta_2,\ \delta)$ send to the designated verifier Cindy and he try to operate the forgery from the condition of the verification i.e. he wants to determine $h\ (= H_1\ (H_2(\beta_1\beta_2) \oplus m))$ such that the condition of verification $(\delta x'_o \sim hx_o,\ \delta \sim h)$ are satisfied. But for obtaining $h$ he must compute $\beta_1\ (= a_c\alpha_1 a_c^{-1})$ and $\beta_2\ (= a_T\alpha_2 a_T^{-1})$ that uses the secret key of the verifier. Hence, he cannot forge the signatures from the condition of verification.

- **Unlinkability:** The two designated verifier do not know each other even then they can verify the signatures individually.
  **Explanation:** The two designated verifiers Cindy and Trevor do not know anything about each other's identity; even then they can verify the signatures individually. Moreover, providing Cindy $\beta_2\ (= a_T\alpha_2 a_T^{-1})$ and Trevor $\beta_1\ (= a_c\alpha_1 a_c^{-1})$ with the signatures does not reveal the identity of the other designated verifier.

## 6. Proposed Designated Verifier Proxy Signature Scheme With Delegation By Warrant

In this section we propose our designated verifier proxy signature scheme. We have added the concept of proxy in the scheme proposed in section 4.

- **Key Generation:** Same as section 3.

- **Proxy key Generation:** The signer Alice chooses a message '$m$', a warrant $m_w$ and a braid $z_o \in LB_l$ and computes $t_o = a_o z_o a_o^{-1}$ . She sends $\sigma = (m_w,\ z_o,\ t_o)$ to the proxy signer Bob. Bob checks if $t_o x'_o \sim z_o x_o.$ If yes, then Bob computes the proxy key $PK = a_p t_o a_p^{-1}$

- **Proxy signature generation:** The proxy signer Bob chooses a braid $b \in LB_l$ and computes $\alpha = bx_c b^{-1}$, $\beta = b x'_c b^{-1}$, $h = H_1 (H_2(\beta) \oplus m_w)$, $\gamma = bhb^{-1}$, $\delta = bx_p b^{-1}$, $\theta = b a_p^{-1} (PK) a_p b^{-1}$. Sends $\sigma' = (m_w, \alpha, \gamma, \delta, \theta, t_o)$ to Cindy.

- **Proxy Signature Verification:** Cindy on receiving $\sigma'$ computes $\beta = a_c \alpha a_c^{-1}$, $h = H_1(H_2(\beta) \oplus m_w))$ and accepts the signature if and only if $\gamma \sim h$, $\delta \sim x_p$, $\gamma\delta \sim hx_p$, $\gamma\theta \sim ht_o$ holds.

- **Correctness:** The following relation give the correctness of the verification equation:
  - $\gamma = bhb^{-1}$

    i.e. $\gamma \sim h$
  - $\delta = bx_p b^{-1}$

    i.e. $\delta \sim x_p$
  - $\gamma\delta = (bhb^{-1})(bxpb^{-1})$

    $= b(hxp)b^{-1}$

    i.e. $\gamma\delta \sim hxp$
  - $\gamma\theta = (bhb^{-1})( b a_p^{-1} (PK) a_p b^{-1})$

    $= (bhb^{-1})( b a_p^{-1} ( a_p t_o a_p^{-1}) a_p b^{-1})$

    $= (bhb^{-1})(bt_o b^{-1} )$

    $= b(ht_o)b^{-1}.$

    i.e. $\gamma\theta \sim ht_o$

- **Applications:**

Consider the scenario of online internet shopping where a vendor Bob is selling goods, produced by Alice. A customer Cindy wishes to buy any of products 'P' like books, music CDs and movies etc. Since Cindy does not completely trust Bob for his products so she needs a digitally signed receipt from Bob so that she can check the originality, authenticity and legality of the product 'P'. Moreover, Cindy also expects that the receipt should not only bounds with the identity of Bob but also that of goods producer Alice. With such a receipt, Cindy will be convinced that goods are produced only by Alice and are being sold by Bob. At the same time Alice and Bob wants that validity of Cindy's receipt can only be verified by Cindy herself and she should not be able to illegally distribute this digitally signed receipt to others. In this situation, Strong Designated Verifier Proxy Signature Schemes can be used to produce digital receipt instead of ordinary signatures, as proposed here.

- **Security Analysis:** In this section we discuss the security aspects of our scheme

➢ **Proxy protected:** Only the proxy signer can sign the message.
**Explanation:** The construction of proxy key involves the secret key of the proxy signer. So, no one other than the proxy signer can construct the proxy key. Moreover, the warrant $m_w$ is attached with the signature that contains the identity of the proxy signer. Also, the proxy key involves a secret braid chosen by the original signer. So, even if one knows the secret key of both the original signer and the proxy signer he cannot construct the valid proxy key.

➢ **Strongness:** Only the designated verifier can check the validity of the signatures.
**Explanation:** The verification of the designated signatures involves the secret key of the designated verifier. So, only the designated verifier can check the validity of the signatures.

➢ **Secrecy of the proxy key:** This is same as proposed in section 3.

## 7. Proposed Bi-Designated Verifier Proxy Signature Scheme With Delegation By Warrant

Now, in this section we add one more designated verifier to the scheme proposed in section 6 to form a new concept of bi-designated verifier signature scheme over braid groups.

- **Key Generation:** Same as section 3.

- **Proxy key Generation:** The signer Alice chooses a message '*m*', a warrant $m_w$ on message '*m*' and a braid $z_o \in LB_l$ and computes $t_o = a_o z_o a_o^{-1}$ and sends *(*$m_w$*,* $z_o$, $t_o$) to the proxy signer Bob. Bob checks $t_o x'_o \sim z_o x_o$. If this holds then computes the proxy key $PK = a_p t_o a_p^{-1}$ .

- **Signature Generation:** The signer Alice chooses a message '*m*' and a braid $b \in LB_l$ and computes $\alpha_1 = bx_c b^{-1}$, $\beta_1 = b x'_c b^{-1}$, $\alpha_2 = bx_T b^{-1}$, $\beta_2 = b x'_T b^{-1}$
  $h = H_1(H_2(\beta_1 \beta_2) \oplus m_w)$, $\gamma = bhb^{-1}$, $\delta = bx_p b^{-1}$, $\theta = b a_p^{-1} (PK) a_p b^{-1}$
  Sends $\sigma_1 = (m, \alpha_1, \beta_2, \gamma, \delta, \theta, m_w)$ to the designated verifier Cindy and $\sigma_2 = (m, \alpha_2, \beta_1, \gamma, \delta, \theta, m_w)$ as the signature on message '*m*' to the designated verifier Trevor through a secure channel.

- **Signature Verification:** Cindy on receiving the signatures computes $\beta_1 = a_c \alpha_1 a_c^{-1}$, $h = H_1(H_2(\beta_1 \beta_2) \oplus m)$, and accepts the signature if and only if $\gamma \sim h$, $\delta \sim x_p$, $\gamma\delta \sim hx_p$, $\gamma\theta \sim ht_o$ holds.
  Similarly, Trevor computes $\beta_2 = a_T \alpha_2 a_T^{-1}$ $h = H_1(H_2(\beta_1 \beta_2) \oplus m)$, and accepts the signature if and only if $\gamma \sim h$, $\delta \sim x_p$, $\gamma\delta \sim hx_p$, $\gamma\theta \sim ht_o$ holds.

- **Correctness:** The following equations gives the correctness of the proposed scheme
  - $\gamma = bhb^{-1}$ i.e. $\gamma \sim h$
  - $\delta = bx_p b^{-1}$ i.e. $\delta \sim x_p$
  - $\gamma\delta = (bhb^{-1})(bxpb^{-1}) = b(hxp)b^{-1}$
    i.e. $\gamma\delta \sim hxp$
  - $\gamma\theta = (bhb^{-1})( b a_p^{-1} (PK) a_p b^{-1})$
    $= (bhb^{-1})( b a_p^{-1} ( a_p t_o a_p^{-1}) a_p b^{-1})$
    $= (bhb^{-1})(bt_o b^{-1} )$
    $= b(ht_o)b^{-1}$
    i.e. $\gamma\theta \sim ht_o$

- **Applications:**

A corporate manager Alice will have vacations for one or two weeks. However, in his absence some current business proposals need to be discussed with the clients. Assistant Bob, as the representative of Alice is assigned proxy signing powers to negotiate with a single business proposal with two different customers Cindy and Trevor in this period and to sign the contract with the person who satisfies their conditions. During this procedure some intermediate documents will be produced for authentication purpose using digital signatures. To protect the confidentiality and authenticity of those documents it may be highly expected that the corresponding signatures could be validated only by the designated receiver. Moreover, they should not be able to convince any third party about these facts. In such cases, Strong Bi-Designated Verifier Proxy Signature Schemes as proposed in chapter 6 could be utilized.

- **Security Analysis:** In this section we discuss the security aspects of our scheme

➢ **Proxy protected:** Only the proxy signer can sign the message.
**Explanation:** Same as the section 4 scheme.

➢ **Strongness:** Only the two designated verifiers can check the validity of the signatures.
**Explanation:** The verification of the designated signatures involves the secret key of the designated verifiers i.e. if Cindy wants to check the validity of the signatures then he has to use his secret key. So, only the designated verifier can check the validity of the signatures.

➢ **Secrecy of the proxy key**: This is same as proposed in section 3.

## 8. Conclusion

In this paper we have proposed a proxy signature scheme with delegation by warrant using conjugacy search problem over the braid groups. Firstly, we have proposed our proxy signature scheme that includes the warrant with the signatures that restricts the proxy signer to create the valid proxy signatures for a certain period of time. We then discuss designated verifier and bi-designated verifier signature schemes. Finally, we added these two concepts of proxy signatures and designated verifier signatures and bi-designated verifier signatures. The security of our schemes is based on the conjugacy problem of braid groups. To the best of our knowledge these are first signature schemes of this type defined over braid groups.